\documentclass{article}
\usepackage{spconf,amsmath,graphicx,mymathdefs}
\usepackage{color}
\usepackage{enumerate}
\usepackage[none]{hyphenat}

\newtheorem{theorem}{Theorem}
\newcounter{remark}
\def\remark{\addtocounter{remark}{1}\emph{Remark~\theremark}. }
\def\C#1x#2{\setC^{#1\times #2}}
\def\CN{${\mathcal C\mathcal N}(0,1)$}
\def\Pd{P_\text{d}}
\def\hH{\Hat H}
\def\tH{\Tilde H}
\def\Yp{Y_\text{p}}

\def\thref#1{Theorem.~\ref{#1}}

\makeatletter
\renewcommand{\section}{
  \@startsection {section}{1}{\z@ }%
  {-2.5ex\@plus -1ex \@minus -.2ex}%
  {0.5ex \@plus .2ex}%
  {\large \bfseries}%
}
\renewcommand{\subsection}{
   \@startsection {subsection}{2}{\z@ }%
   {-1.5ex\@plus -.6ex \@minus -.2ex}%
   {0.5ex \@plus .2ex}%
   {\bfseries}%
}
\makeatother

\normalsize

\title{Performance of Uplink Multiuser Massive MIMO Systems}

\name{Zhengdao Wang}

\address{Dept. of Electrical \& Computer Engineering, Iowa State University.
Email: zhengdao@iastate.edu}

\begin{document}
\maketitle
\begin{abstract}
We study the performance of uplink transmission in a large-scale (massive)
MIMO system, where all the transmitters have single antennas and the receiver
(base station) has a large number of antennas. Specifically, we analyze
achievable degrees of freedom of the system without assuming channel state
information at the receiver. Also, we quantify the amount of power saving that
is possible with increasing number of receive antennas.
\end{abstract}

\section{Introduction}

Massive multiple-input multiple-output (MIMO) systems are a type of cellular
communication where the base station is equipped with a large number of
antennas. The base station serves multiple mobile stations that are usually
equipped with a small number of antennas, typically one. Massive MIMO holds
good potentials for improving future communication system performance. There
are several challenges with designing such massive MIMO systems, including
e.g., channel state information (CSI) acquisition, base station received
signal processing, downlink beamforming with imperfect CSI, etc. For
multi-cell system, pilot contamination and inter-cell interference also need
to be dealt with. There is already a body of results in the literature about
the analysis and design of large MIMO systems; see e.g., the overview article
\cite{rpll13} and references there in.

To reveal the potential that is possible with massive MIMO systems, it is
important to quantify the achievable performance of such systems in realistic
scenarios. For example, it is too optimistic to assume that perfect CSI can be
acquired at the base station in the uplink, because such acquisition takes
time, energy, and channel estimation error will always exist. For the
downlink, in order to perform effective beamforming, CSI is again needed,
which needs to be either estimated by the mobile stations and then fed back to
the base station, which is a non-trivial task, or, acquired by the base
station by exploiting channel reciprocity in a time-division duplexing setup.

In this paper, we are interested in performance of the \emph{uplink}
transmission in a \emph{single-cell} system. In particular, we ask what rates
can be achieved in the uplink by the mobile users if we assume realistic
channel estimation at the base station. Similar analysis has been performed in
\cite{nglm13}, but the analysis therein assumes equal power transmission
during the channel training phase and the data transmission phase. Also the
training duration has not be optimized, and the effect of channel coherence
interval was not incorporated.

Our main analysis framework is similar to that of \cite{haho03}. The system we
are analyzing can be viewed as a point-to-point MIMO channel if the mobile
stations are allowed to fully cooperate. Therefore, the rates obtained in
\cite{haho03}, and the stronger result on non-coherent MIMO channel capacity
in \cite{zhts02} can serve as upper bound for the system sum rate.
Furthermore, the training strategy as optimized in \cite{haho03} involves
orthogonal signaling from the multiple transmit antennas and therefore is
directly applicable to the multi-user single-cell system of interest.

For a system with $K$ mobile users, $M$ base station antennas, and block
fading channel with coherence interval $T$, we derive achievable rate using
linear channel estimation and linear base station (front-end) processing; see
\secref{sec.ach}. The total degrees of freedom (DoF), to be defined more
precisely later, is quantified in \thref{th.dof}. We also quantify the needed
transmission power for achieving a given rate, when $M\gg 1$, which is an
refinement of the corresponding result in \cite{nglm13}.

\section{System Model}

Notation: We use $A^\dag$ to denote the Hermitian transpose of a matrix $A$,
$I_K$ to denote a \size KxK identity matrix, $\setC$ to denote the complex
number set, $\lfloor\cdot \rfloor$ to denote the integer floor operation, \iid
to denote ``independent and identically distributed'', and \CN to denote
circularly symmetric complex Gaussian distribution with zero mean and unit
variance.

We consider a single-cell uplink system, where there are $K$ mobile users and
one base station. Each user has one transmit antenna, and the base station has
$M$ receive antennas. The received signal at the base station is
\begin{equation}
  y = Hs + n
\end{equation}
where $H\in \C MxK$ is the channel matrix, $s\in \C Kx1$ is the transmitted
signals from all the $K$ users; $n\in \C Mx1$ is the additive noise, $y\in \C
Mx1$ is the received signal. We make the following assumptions:

A1) The channel is block fading such that within a \emph{coherence interval}
of $T$ channel uses or time slots, the channel remains constant. The entries
of $H$ are \iid and taken from \CN. The channel changes independently from
block to block. The CSI is neither available at the transmitters nor at the
receiver.

A2) Entries of the noise vector $n$ are \iid and from \CN. Noises in different
channel uses are independent.

A3) The average transmit power per user is $P$. So within a coherence interval
the total transmitted energy is $PT$. We do not impose a peak power
constraint.

In summary, the system has four parameters, $(M,K,T,P)$. We will allow the
system to operate in the ergodic regime, so coding and decoding can
occur over multiple coherent intervals.

\section{Achievable rates} \label{sec.ach}

We assume that $K\le M$ and $K<T$ in this section. To derive the achievable
rates for the users, we use a well-known scheme that consists of two phases
(see e.g., \cite{haho03}):

\noindent \emph{Training Phase}. This phase consists of $K$ time intervals.
The training signal transmitted by the users can be represented by a \size KxK
matrix $\Phi$ such that $\Phi\Phi^\dag=E I_K$, where $E$ is the total training
energy per user per coherent interval.

\noindent \emph{Data Transmission Phase}. Information-bearing symbols are
transmitted by the users in the remaining $T-K$ time intervals. The average
energy per symbol per user is $\Pd=(PT-E)/(T-K)$.

\subsection{Channel estimation}

In the training phase, we will choose $\Phi=\sqrt{E}I_K$ for simplicity. Other
scaled unitary matrix can also be used without affecting the achievable rate.
Note that the transmission power is allowed to vary from the training phase to
the data transmission phase. Also, setting the training period equal to the
total number of transmit antennas possesses certain optimality as derived in
\cite{haho03}. With our choice of $\Phi$, the received signal $\Yp \in \C MxK$
during the training phase can be written as
\begin{equation}
  \Yp=H\Phi + N = \sqrt{E}H+N
\end{equation}
where $N\in \C MxK$ is the additive noise. The equation describes
$M\times K$ independent identities, one for each channel coefficient. The
(linear) minimum mean-squared error (MMSE) estimate for the channel $H$ is
given by
\begin{equation}
  \hH = \frac{\sqrt E}{E+1}  \Yp = \frac{E}{E+1}H + \frac{\sqrt E}{E+1} N.
\end{equation}
The channel estimation error is defined as
\begin{equation}
\tH=H-\hH=\frac 1{E+1}H - \frac{\sqrt E}{E+1}N.
\end{equation}
It is well known and easy to verify that the elements of $\hH$ are \iid
complex Gaussian with zero mean and variance
\def\sigmah{\sigma_{\hH}^2}
\begin{equation}
  \sigmah = \frac{E}{E+1},
\end{equation}
and the elements of $\tH$ are \iid complex Gaussian with zero mean and
variance
\def\sigmat{\sigma_{\tH}^2}
\begin{equation}
  \sigmat = \frac{1}{E+1}.
\end{equation}
Moreover, $\hH$ and $\tH$ are in general uncorrelated as a property of linear
MMSE estimator, and in this case independent thanks to the Gaussian
assumptions.

\subsection{Equivalent channel} Once the channel is estimated, the base
station has $\hH$ and will decode the users' information using $\hH$. We can
write the received signal as
\begin{equation} \label{eq.equiv}
  y=\hH s + \tH s + n  := \hH s + v
\end{equation}
where $v\bydef \tH s + n$ is the new equivalent noise containing actual noise
$n$ and self interference $\tH s$ caused by inaccurate channel estimation.
Assuming that each element of $s$ has variance $\Pd$ during the data
transmission phase, and there is no cooperation among the users, the variance
of each component of $v$ is
\def\sigmav{\sigma_v^2}
\begin{equation}
\sigmav=\frac{K\Pd}{E+1}+1.
\end{equation}
If we replace $v$ with a zero-mean complex Gaussian noise with equal variance
$\sigmav$, but independent of $s$, then the system described in
\eqref{eq.equiv} can be viewed as MIMO system with perfect CSI at the
receiver, and equivalent signal to noise ratio (SNR)
\begin{equation}\label{eq.eff}
 \rho \bydef \frac {\Pd\sigmah}{\sigmav} = \frac{\Pd E}{K\Pd + E +1}
 =\frac{\Pd}{1+\frac{K\Pd+1}E}.
\end{equation}
The SNR is the signal power from a single transmitter per receive antenna
divided by the noise variance per receive antenna. It is a standard argument
that a noise equivalent to $v$ but assumed independent of $s$ is ``worse'';
see e.g., \cite{haho03}. As a result, the derived rate based on such
assumption is achievable. In the following, for notational brevity, we assume
that $v$ in \eqref{eq.equiv} is independent of $s$ without introducing a new
symbol to represent the equivalent \emph{independent} noise.

Note that the effective SNR $\rho$ is the actual SNR $\Pd$ divided by a loss
factor $1+(K\Pd+1)/E$. The loss factor can be made small if the energy $E$
used in the training phase is large.

\subsection{Energy splitting optimization} \label{sec.split}

The energy in the training phase can be optimized to maximize the effective
SNR $\rho$ in \eqref{eq.eff}, as has been done in \cite[Theorem~2]{haho03}. We
adapt the result below for our case because it is relevant to our discussion.
Specifically, let $\alpha\bydef E/(PT)$ denote the percentage of energy
devoted to training within one coherent interval. Define an auxiliary variable
when $T\ne 2K$:
\begin{equation}
  \gamma\bydef \frac{(1+PT)(T-K)}{PT(T-2K)}
\end{equation}
which is positive if $T>2K$ and negative if $T<2K$. The optimal value for
$\alpha$ that maximizes $\rho$ is given as follows:
\begin{equation}
  \alpha=\begin{cases}
  \gamma-\sqrt{\gamma(\gamma-1)}, & T>2K \\
  \frac 12, & T=2K \\
  \gamma+\sqrt{\gamma(\gamma-1)}, & T<2K
  \end{cases}
\end{equation}
The maximized effective SNR $\rho$ is given as
\begin{equation}
  \rho=\begin{cases}
\frac{PT}{T-2K}(\sqrt{\gamma}-\sqrt{\gamma-1})^2, & T>2K \\
\frac{(PT)^2}{2T(1+PT)}, & T=2K \\
\frac{PT}{2K-T}(\sqrt{-\gamma}-\sqrt{-\gamma+1})^2, & T>2K
\end{cases}
\end{equation}
At high SNR ($P\gg 1$), the optimal values are
\begin{equation}
  \alpha=\frac{\sqrt{T-K}}{\sqrt{T-K}+\sqrt{K}}, \quad
  \rho=\frac{T}{(\sqrt{T-K}+\sqrt{K})^2}P.
\end{equation}
At low SNR ($P\ll 1$), the optimal values are
\begin{equation}\label{eq.lowsnr}
  \alpha=\frac 12, \quad
  \rho=\frac{(PT)^2}{4(T-K)}.
\end{equation}

\subsection{Achievable rates}

Given the channel model \eqref{eq.equiv}, linear processing can be applied to
$y$ to recover $s$, as in e.g., \cite{nglm13}. Let $A\in \C KxM$ denote the
linear processing matrix. The processed signal is
\def\hs{\Hat s}
\begin{equation}
  \hs \bydef A y = A\hH s + A v.
\end{equation}
The Maximum Ratio Combining (MRC) processing is obtained by setting
$A=\hH^\dag$. The Zero-Forcing (ZF) processing is obtained by setting
$A=(\hH^\dag\hH)^{-1}\hH^\dag$.

Based on the equivalent channel model, viewed as a multi-user MIMO systems
with perfect receiver CSI and equivalent SNR $\rho$, the achievable rates
lower bounds derived in \cite[Propositions~2 and~3]{nglm13} can then be
applied. Specifically, for MRC the following ergodic rate per user is
achievable:
\begin{equation}\label{eq.ratemrc}
  R^\text{(MRC)} \bydef\left (1 - \frac KT\right) \log_2\left(1 +
  \frac{\rho(M-1)}{\rho(K-1)+1} \right).
\end{equation}
For ZF, the following rate per user is achievable:
\begin{equation} \label{eq.ratezf}
  R^\text{(ZF)} \bydef\left (1 - \frac KT\right) \log_2\left[1 +
  \rho(M-K) \right].
\end{equation}
Note that the factor $(1-\frac KT)$ is due to the fact that during one
coherence interval of length $T$, $K$ time slots have been used for the
training purpose. The number of data transmission slots is $T-K$, and the
achieved rate needs to be averaged over $T$ channel uses.

\section{Degrees of freedom}

We define the DoF of the system as
\begin{equation}
  d(M, K, T)\bydef \sup \lim_{P\to\infty} \frac{R^{(\text{total})}(P)}{\log_2 (P)}
\end{equation}
where the supremum is taken over the totality of all reliable communication
schemes for the system, and $R^{(\text{total})}$ denotes the sum rate of the
$K$ users under the power constraint $P$. We may also speak of the (achieved)
degree of freedom of one user for a particular achievability scheme, which is
the achieved rate of the user normalized by $\log_2(P)$ in the limit of $P\to
\infty$. The DoF measures the multiplexing gain offered by the system when
compared to a reference point-to-point single-antenna communication link, in
the high SNR regime.

\begin{theorem} \label{th.dof}
For an $(M, K, T)$ MIMO uplink system with $M$ receive antennas, $K$ users,
and coherence interval $T$, the total DoF of the system is
\begin{equation}
  d(M,K,T)=K^*\left(1-\frac {K^*}{T}\right).
\end{equation}
\end{theorem}
where $K^*\bydef \min(M, K, \lfloor T/2\rfloor)$. \QED

\emph{Proof:} To prove the converse, we observe that if we allow the $K$
transmitters to cooperate, then the system is a point-to-point MIMO system
with $K$ transmit antennas, $M$ receive antennas, and with no CSI at the
receiver. The DoF of this channel has been quantified in \cite{zhts02}, in
the same form as in the theorem. Without cooperation, the users can at most
achieve a rate as high as in the cooperation case.

To prove the achievability, we first look at the case $K^*<M$. In this case,
we note that if we allow only $K^*$ users to transmit, and let the remaining
users be silent, then using the achievability scheme describe in
\secref{sec.ach}, each of the $K^*$ users can achieve a rate per user using
the zero-forcing receiver given as follows (cf.~\eqref{eq.ratezf})
\begin{equation}
  \left (1 - \frac {K^*}T\right) \log_2\left[1 +
  \rho(M-K^*) \right].
\end{equation}
Note that the condition $K^*<M$ is needed. If we choose $E=KP$ and $\Pd=P$,
then the effective SNR in \eqref{eq.eff} becomes
\begin{equation} \label{eq.highsnr}
  \rho=\frac{P}{1+\frac{KP+1}{KP}}.
\end{equation}
It can be seen that as $P\to\infty$, $\log(\rho)/\log(P)\to 1$ and a DoF per
user of $(1-K^*/T)$ is achieved. The total achieved DoF is therefore
$K^*(1-K^*/T)$. Although better energy splitting is possible, as in
\secref{sec.split}, it will not improve the DoF.

When $K^*=M$, the case is more subtle. In this case the zero-forcing receive
is no longer sufficient. In fact, even the optimal linear processing, which is
the MMSE receiver \cite[eq.~(31)]{nglm13}, is not sufficient. The
insufficiency can be established by using the results in
\cite[Sec.IV.C]{gasc98} to show that as $P\to\infty$, the effective SNR at the
output of MMSE receiver has a limit distribution that is independent of SNR.
We skip the details here.

Instead, we notice that the equivalent channel \eqref{eq.equiv} has SNR given
by \eqref{eq.highsnr}, which for $KP>1$ is greater than $P/3$. So, the MIMO
system can be viewed as a Multiple Access Channel (MAC) with $K^*$
single-antenna transmitters, and one receiver with $M$ receive antennas.
Perfect CSI is known at the receiver, and the SNR between $P/3$ and $P$. Using
the MAC capacity region result \cite[Theorem~14.3.1]{coth91},
\cite[Sec.~10.2.1]{tse05}, it can be shown that a total DoF of $K^*$ can be
achieved over $T-K^*$ the time slots.\QED

\remark The DoF is the same as that of a point-to-point MIMO channel with $K$
transmit antennas and $M$ receive antennas without transmit- or receive-side
CSI \cite{zhts02}. This is a bit surprising because optimal signaling over
non-coherent MIMO channel generally requires cooperation among the transmit
antennas. It turns out that as far as DoF is concerned, transmit antenna
cooperation is not necessary. This is the new twist compared to the
point-to-point case.

\remark It can be seen from the achievability proof that for $M>K$, which is
generally applicable for ``massive'' MIMO systems, zero-forcing at the base
station is sufficient for achieving the optimal DoF. However, the MRC is not
sufficient because $\rho$ shows up both in the numerator and denominator of
\eqref{eq.ratemrc}. So as $\rho\to\infty$, the achieved rate is limited. This
is due to the interference among the users.

\remark For the case $K^*=M$, non-linear decoding such as successive
interference cancellation is needed.

\remark When $T$ is large, a per-user DoF close to 1 is achievable, as long as
$K\le M$.

\remark When $M$ is larger than $K^*$, increasing $M$ further has no effect on
the DoF. However, it is clear that more receive antennas is useful because
more energy is collected by additional antennas. We will discuss the benefit
of energy savings in the next section.

\section{Discussion}

\subsection{Power savings for fixed rate}

As more antennas are added to the base station, more energy can be collected.
Therefore, it is possible that less energy is needed to be transmitted from
the mobile stations. When there is perfect CSI at the base station, it has
been shown in \cite{nglm13} that the transmission power can be reduced by a factor
$1/M$ to maintain the same rate, compared to a single-user single-antenna
system.

When there is no CSI at the receiver, however, it was observed in
\cite{nglm13} that the power savings factor is $1/\sqrt{M}$ instead of $1/M$.
In the following we do a slightly finer analysis of the effected power savings
when $M$ is large, assuming the training phase has been optimized as in
\secref{sec.ach}.

Consider $M\gg K> 1$. Because the received power is linearly proportional to
$M$, the transmitted power can be smaller when $M$ is larger. When $M\gg 1$,
the system is operating in power-limited regime. It can be seen from
\eqref{eq.ratemrc} and \eqref{eq.ratezf} that when $\rho$ is small, MRC
performs better than ZF, which has been previously observed, e.g.,
\cite{nglm13}. On the other hand, in the low-SNR regime the difference between
them is a constant factor $(M-1)/(M-K)$ in the SNR term within the logarithmic
functions in \eqref{eq.ratemrc} and \eqref{eq.ratezf}. The difference becomes
negligible when $M$ is large. Using either result, and the effective SNR in
\eqref{eq.lowsnr}, we see that if we fix the per-user rate at
$R=(1-K/T)\log_2(1+\rho_0)$, then the required power $P$ can be found by
setting $\rho M=\rho_0$, resulting in
\begin{equation}\label{eq.saving}
  P=\sqrt{\frac{4\rho_0 (T-K)}{MT^2}} + o\left(\frac1{\sqrt M} \right)
\end{equation}
It is interesting to note that increasing $T$ has a similar effect as
increasing $M$ on the required transmission power, reducing the power by
$1/\sqrt{M}$ or $1/\sqrt{T}$. The reason is the if $T$ is increased, then the
energy that can be expended on training is increased, improving the quality of
channel estimation. On the other hand, for \eqref{eq.saving} to be applicable,
we need $M\gg K$.

\subsection{Peak power limited case}

If the peak power is limited rather than the average power, then our DoF
result still holds because the achievability proof actually uses equal power
in the training and data transmission phases. The power savings discussion in
the previous subsection still applies, because the system is limited by the
total amount of energy available, and not how the energy is expended. In the
regime where the SNR is neither very high or very low, the peak power
constraint will affect the rate. A detailed analysis is not included here.

\subsection{MMSE and optimal processing}

If MMSE processing is used at the base station, then the performance can be
improved compared to MRC and ZF. However, at low SNR, MRC is near optimal and
at high SNR, ZF is near optimal. So MMSE processing will not change the nature
of the results that we have obtained, although a slightly higher rate is
possible. It is also possible to analyze the achievable rate with optimal
non-linear processing, using known MAC capacity region results.

\clearpage
\bibliographystyle{IEEE}
\bibliography{refs}

\begin{thebibliography}{1}

\bibitem{coth91}
T.~M. Cover and J.~A. Thomas,
\newblock {\em Elements of Information Theory},
\newblock John Wiley \& Sons, Inc., 1991.

\bibitem{gasc98}
H.~Gao, P.~J. Smith, and M.~V. Clark,
\newblock ``Theoretical reliability of {MMSE} linear diversity combining in
  rayleigh-fading additive interference channels,''
\newblock {\em {IEEE} Trans. Commun.}, vol.~46, no.~5, pp.~666--672, May 1998.

\bibitem{haho03}
B.~Hassibi and B.~M. Hochwald,
\newblock ``How much training is needed in multiple-antenna wireless links?''
\newblock {\em {IEEE} Trans. Info. Theory}, vol.~49, no.~4, pp.~951--963, Apr.
  2003.

\bibitem{nglm13}
H.~Q. Ngo, E.~G. Larsson, and T.~L. Marzetta,
\newblock ``Energy and spectral efficiency of very large multiuser {MIMO}
  systems,''
\newblock {\em {IEEE} Trans. Commun.}, vol.~61, no.~4, pp.~1436--1449, Apr.
  2013.

\bibitem{rpll13}
F.~Rusek, D.~Persson, B.~K. Lau, E.~G. Larsson, T.~L. Marzetta, O.~Edfors, and
  F.~Tufvesson,
\newblock ``Scaling up {MIMO:} opportunities and challenges with very large
  arrays,''
\newblock {\em IEEE Signal Processing Magazine}, vol.~30, no.~1, pp.~40--60,
  Jan. 2013.

\bibitem{tse05}
D.~Tse and P.~Viswanath,
\newblock {\em Fundamentals of Wireless Communication},
\newblock Cambridge University Press, 2005.

\bibitem{zhts02}
L.~Zheng and D.~Tse,
\newblock ``Communicating on the {Grassmann} manifold: A geometric approach to
  the non-coherent multiple antenna channel,''
\newblock {\em {IEEE} Trans. Info. Theory}, vol.~48, no.~2, pp.~359--383, Feb.
  2002.

\end{thebibliography}
\end{document}